# Engineering high-Q superconducting tantalum microwave coplanar waveguide resonators for compact coherent quantum circuits

Shima Poorgholam-Khanjari, Valentino Seferai, Paniz Foshat, Calum Rose, Hua Feng, Robert H. Hadfield, Martin Weides, and Kaveh Delfanazari [*]

*Electronics and Nanoscale Engineering Division, James Watt School of Engineering, University of Glasgow, Glasgow, UK*
[*]Corresponding author: kaveh.delfanazari@glasgow.ac.uk, Dated: 20/12/2024

Tantalum (Ta) has recently received considerable attention in manufacturing robust superconducting quantum circuits. Ta offers low microwave loss, high kinetic inductance compared to aluminium (Al) and niobium (Nb), and good compatibility with complementary metal-oxide-semiconductor (CMOS) technology, which is essential for quantum computing applications. Here, we demonstrate the fabrication engineering of thickness-dependent high quality factor (high-$Q_i$) Ta superconducting microwave coplanar waveguide resonators. All films are deposited on high-resistivity silicon substrates at room temperature without additional substrate heating. Before Ta deposition, a niobium (Nb) seed layer is used to ensure a body-centred cubic lattice ($\alpha$-Ta) formation. We further engineer the kinetic inductance ($L_K$) resonators by varying Ta film thicknesses. High $L_K$ is a key advantage for applications because it facilitates the realisation of high-impedance, compact quantum circuits with enhanced coupling to qubits. The maximum internal quality factor $Q_i$ of $\sim 3.6\times 10^6$ is achieved at the high power regime for 100 nm Ta, while the highest kinetic inductance is obtained 0.6 pH/sq for the thinnest film, which is 40 nm. This combination of high $Q_i$ and high $L_K$ highlights the potential of Ta microwave circuits for high-fidelity operations of compact quantum circuits.

Superconducting microwave coplanar waveguide (CPW) resonators are one of the fundamental components of circuit quantum electrodynamics (cQEDs), and quantum computing chips owing to their high-quality factor resonances and the simplicity of the fabrication process. Additionally, their characteristic impedance can be tuned by modifying the gaps' width and the centre conductor's width [1-4]. However, two-level systems (TLSs) and quasi-particle losses significantly affect the performance of superconducting circuits. TLS loss is dominant at low power and temperatures and exists in metal-substrate, metal-air and substrate-air [5-9] interfaces. Recent efforts have been focused on reducing losses to reach higher quality factors [10-16]. The fabrication of high-performance qubits is highly feasible with superconducting films that exhibit low dielectric losses at surfaces and interfaces [17-20]. Enhancing fabrication quality and minimizing interfaces help to decrease the losses [21-23]. High-$Q$ superconducting resonators, which exhibit both improved coherence and enhanced transmission are useful for hybrid [24, 25] and unconventional [26] circuits in emerging quantum applications [27-29]. Tantalum (Ta) is one of the materials that is often used in superconducting circuits and has recently attracted significant attention in the manufacturing of robust superconducting quantum circuits. It offers superior performance due to its low intrinsic losses, relatively high superconducting transition temperature, chemical stability and compatibility with semiconductor manufacturing processes, such as CMOS technology [30-34]. The native oxide layer of Ta is thinner than other materials [35], resulting in lower loss and higher coherence time in superconducting qubits [17, 18]. The potential of body-centred cubic (BCC) lattice ($\alpha$-Ta) film as a superconductor in the development of high-performance, large-scale superconducting quantum circuits could facilitate the realization of practical superconducting quantum computers [21]. It can be grown on heated sapphire substrates without a buffer layer [30, 36, 37], on heated sapphire substrates with a buffer layer [38], or on unheated/heated silicon substrates with/without a seed layer [31-33, 35, 39]. However, sapphire is an insulator and an extremely hard material, making it incompatible with large-scale, conventional CMOS fabrication technologies. The use of advanced integration techniques, such as through silicon vias (TSVs) technology to scale up sapphire substrates for growing $\alpha$-Ta films is challenging. In contrast, silicon substrates are primarily used for large-scale integrated circuits [21, 35].

On the other hand, reducing the thickness of superconducting films increases their kinetic inductance, leading to a shift in the resonance frequencies of superconducting resonators. The field of cQED is increasingly influenced by materials with high kinetic inductance. The unique properties of high kinetic inductance materials can open up new possibilities for quantum information processing and applications in sensing and metrology. Moreover, materials with high kinetic inductance are valuable for enhancing device performance in superconducting electronics, quantum computing, and other cutting-edge technologies. These include microwave detectors [40, 41], parametric amplifiers [42, 43], fluxonium qubits [44, 45], resonators [46], and high-coherence quantum processors [47, 48]. However, thin films serve additional purposes. For example, in superconducting nanowire single-photon detectors (SNSPDs), thin films are used to suppress the critical temperature, which, in turn, suppresses the superconducting energy gap $\Delta$. Suppressing the superconducting gap is beneficial for improving photon detection efficiency [49]. High kinetic inductance materials can manipulate the signals with better control and precision in the circuits.

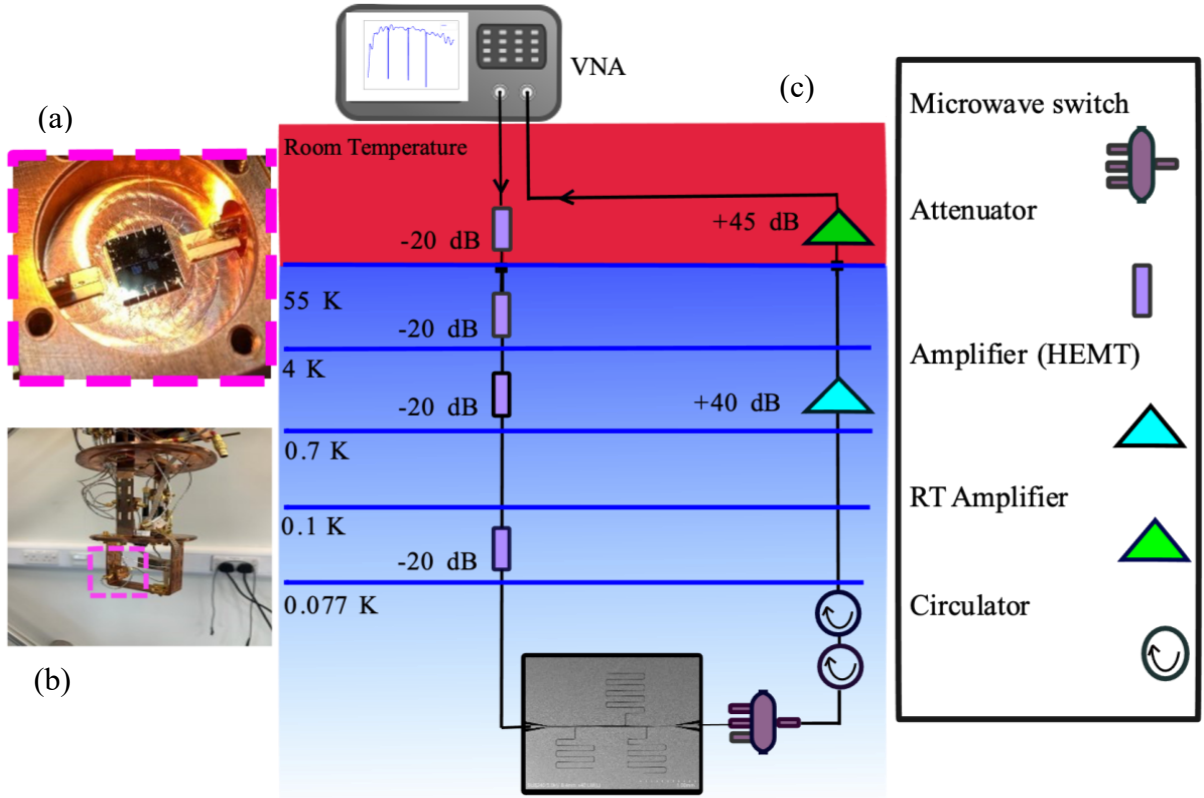

Figure 1. (a) The microwave superconducting coplanar waveguide resonators based on 40 nm thick Ta on silicon chip wire bonded to a copper sample box. (b) The packaged Ta microwave superconducting chip is mounted in the 77 mK stage of a dilution refrigerator. (c) Schematic of the cryogenic setup for sub-Kelvin microwave spectroscopy of the chip.

In this work, we use a silicon chip as the substrate and $\alpha$-Ta is the superconducting material to design and fabricate microwave CPW resonators. Before Ta deposition, we sputtered an Nb seed layer. This process was carried out without heating the substrate, making the approach versatile for a wide range of applications in classical and quantum technologies. We explore the thickness-dependent properties of Ta films with thicknesses of 40 nm, 80 nm and 100 nm with a particular emphasis on the thinnest sample, and demonstrate that thin Ta CPW resonators exhibit a competitive kinetic inductance value compared to previously reported works [30-33, 35, 39]. Following the fabrication, we analyze the effects of temperature and microwave power on the resonators' internal quality factors.

Our design comprises three quarter-wavelength CPW Ta microwave resonators, each featuring a central line width of 4 μm and gaps of 2 μm wide. All resonators are coupled to a common transmission line. Before deposition, a (110) oriented silicon wafer with a resistivity of 20 kΩ·cm and a thickness of 525 μm was cleaned with acetone, isopropanol (IPA), and reverse osmosis (RO) water, respectively, to remove any particles and residues. Then, an MP 600 S Plassys sputter system was used to sputter Ta films. Before Ta deposition, a 5 nm Nb seed layer was sputtered to promote the growth of the Ta $\alpha$-phase. MP 600 S Plassys sputter system is based on DC sputtering principles. We did a pre-sputter process for both Ta (P = 318 W, I = 0.801 A and V = 400 v) and Nb films (P = 228 W, I = 0.8 A and V = 286 v). Then we did $3':30"$ sputtering for Ta with (P = 291 W, I = 0.802 A and V = 360 v) and $45"$ sputtering for Nb (P = 195 W, I = 0.8 A and V = 244.7 v).

Next, the ZEP520A resist [46] was spun on the chip and soft-baked at 180 °C for four minutes. Patterns were then formed using electron-beam (e-beam) lithography, followed by dry etching with an ICP 180 etching tool with CF4/Ar recipe (CF4 = 10 sccm, Ar = 5 sccm with RF power = 10 W and ICP power = 200 W). The resonance frequencies of the resonators were designed to be between 4-8 GHz. Finally, the structures were diced into $5 \times 5$ mm$^2$ chips. One selected chip was wire-bonded to a copper sample box (Fig.1 (a)) and mounted to an Oxford Instruments Triton 200 Dilution Refrigerator (DR) system (Fig.1 (b)) where it was cooled down to a base temperature of $T = 77$ mK. Figure 1. (c) shows the schematic of the DR used for the measurement of the chip. The input signals from the Vector Network Analyzer (VNA) signals are attenuated by 20 dB at room temperature and 60 dB attenuations inside the fridge before reaching the transmission line of the superconducting circuit. Then after passing through the chip, the output signals were amplified by a high-electron mobility transistor (HEMT) low-noise amplifier with a 40 dB gain at the 4 K stage and by a room temperature (RT) amplifier with a 45 dB gain.

Several methods can be used to extract the resonator parameters such as the internal quality factor ($Q_i$), loaded quality factor ($Q_l$), coupling quality factor ($Q_c$), and resonance frequency ($f_r$) from the measurement data obtained using a VNA. The conventional methods are based on either the amplitude or phase of the $S_{21}$ [50, 51], while recent methods utilize the full complex scattering data to achieve a more precise calculation of the resonator's parameters [52-54]. In this work, we employ a notch-type model [53] to extract the resonator parameters:

$$S_{21}^{notch}(f) = \underbrace{a e^{i\alpha} e^{-2\pi i f \tau}}_{A} \times \underbrace{\left(1 - \frac{\left(\frac{Q_l}{|Q_c|}\right) e^{i\varphi}}{1 + 2i Q_l \left(\frac{f}{f_r} - 1\right)}\right)}_{B} \quad (1)$$

Part $B$ of Eq. (1) describes an ideal notch-type resonator, where $f$ represents the probe frequency, $\varphi$ quantifies the impedance mismatch and $|Q_c|$ is the absolute value of the coupling quality factor. Part $A$ of Eq. (1) defines the environment. Amplitude $a$ shows the cable damping effect in $S_{21}$, $\alpha$ a phase shift and $\tau$ represents the electronic delay caused by the length of the cable and the speed of light. We characterized the $T_c$ of different thicknesses of 40 nm, 80 nm and 100 nm Ta films (Fig.2 (a)) by DC measurements. The $T_c$ of the samples were found to be 4.06 K, 4.2 K and 4.49 K, respectively, confirming high-quality superconducting thin film production. Using these critical temperatures and extracting sheet resistances, we could estimate the kinetic inductance ($L_K$) of the Ta samples by using Eq. (2) [46]:

$$L_K \approx \frac{\hbar R_s}{\pi \Delta_0} \qquad (2)$$

Where $\hbar$, is reduced Planck constant, $R_s$ is sheet resistance at normal-state and $\Delta_0$ is the superconducting gap at zero temperature. Assuming $\Delta_0 = 1.76 k_B T_c$ [46, 55], where $k_B$ is the Boltzmann constant, we calculated $L_K \approx 0.6$ (pH/sq), 0.25 (pH/sq), and 0.2 (pH/sq) for 40 nm, 80 nm and 100 nm, respectively. Using conformal mapping techniques, inductance per unit length ($L_l$) and capacitance per unit length ($C_l$) of a CPW resonator can be obtained [56, 57], $L_l = 4.13 \times 10^{-7}$ H/m, $C_l = 1.73 \times 10^{-10}$ F/m and $Z_0 = 49$ Ω. By using $v_{ph} = \frac{1}{\sqrt{C_l (L_m + L_k)}}$ and substituting the $v_{ph} = \frac{\omega_n}{k_n}$, $k_n = \frac{\pi}{2l}$, the $L_K$ (H/m) can be calculated for all thicknesses. Figure 2. (b) shows the calculated $L_K$ (H/m) for 40 nm, 80 nm and 100 nm Ta film thicknesses. It shows that by decreasing the thickness, the $L_K$ increases. For the Ta film, the effective penetration depth can be obtained by $\lambda_{eff} = \lambda_0 \coth(\frac{d}{\lambda_0})$ [38, 58]; where $\lambda_0$ represents the penetration depth of the bulk superconductor which is $\lambda_0 = 150$ nm for Ta [38, 59], and $d$ is the film thickness. Figure 2. (c) shows the calculated penetration depth of fabricated samples. As can be seen, by increasing the thickness, due to a reduction in kinetic inductance the penetration depth decreases. The fundamental resonance frequency ($f_0$) of the $\frac{\lambda}{4}$ resonators can be obtained using Eq. (3) [36, 60, 61] and higher-order harmonics occur at $3f_0$, $5f_0$, $7f_0$, etc:

$$f_0 = \frac{c}{4l\sqrt{\varepsilon_{eff}}} \qquad (3)$$

Where $c$ is the speed of light ($c \approx 3 \times 10^8$ m/s), $l$ is the length of the resonator, and $\varepsilon_{eff} = \frac{\varepsilon_r + 1}{2}$ is the effective permittivity with $\varepsilon_r$ is the relative permittivity of the substrate.

Figure 3. (a) shows the frequency spectrum of a Ta microwave CPW chip with three resonators, measured at $T = 77$ mK. The spectrum reveals three resonance frequencies, $f_r$, within the 3-5 GHz range. The fitted and measured amplitude and circle fit for $f_r = 3.654$ GHz at high power (VNA power = 0 dBm) and extracted $Q_l = 4872$ and $Q_c = 4897$ are shown in Fig.3(b).

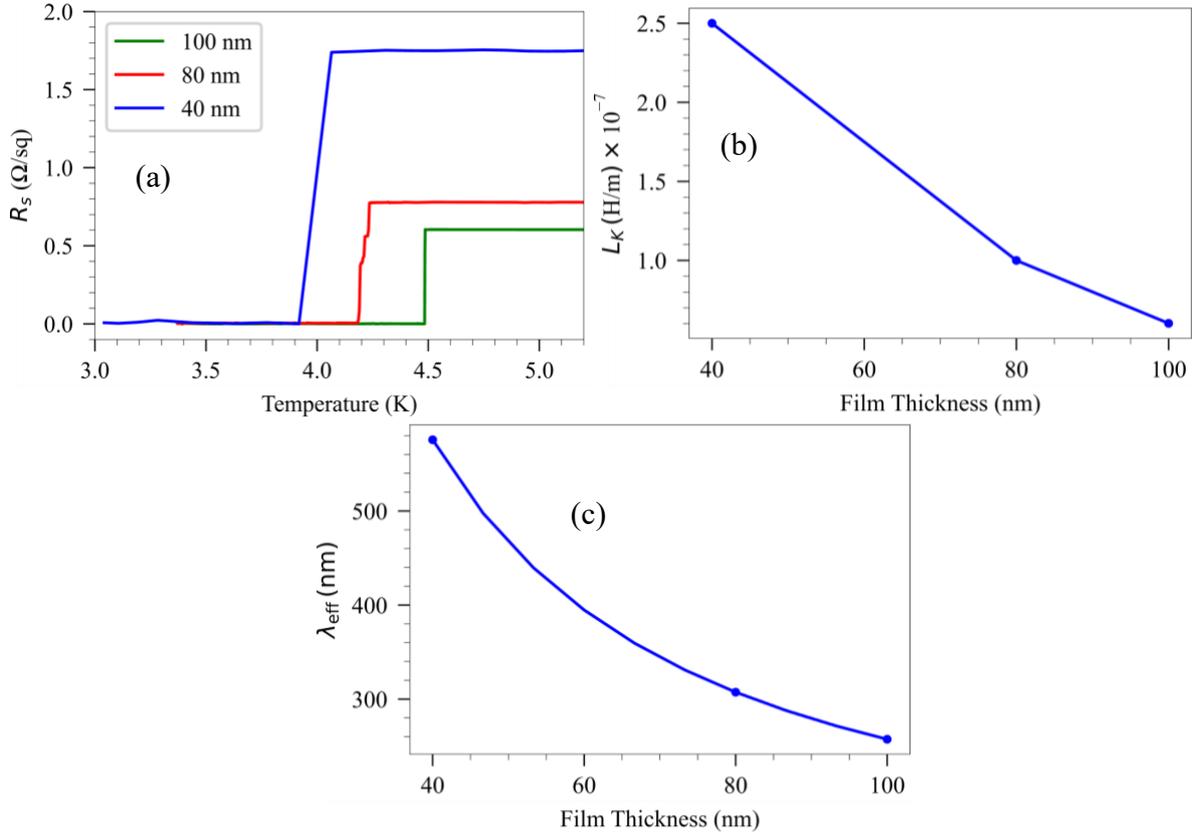

Figure 2. (a) Measured sheet resistance $R_s$ and (b) Calculated $L_K$ (H/m) for Ta with 40 nm, 80 nm and 100 nm thicknesses. (c) Calculated penetration depth as a function of film thickness d.

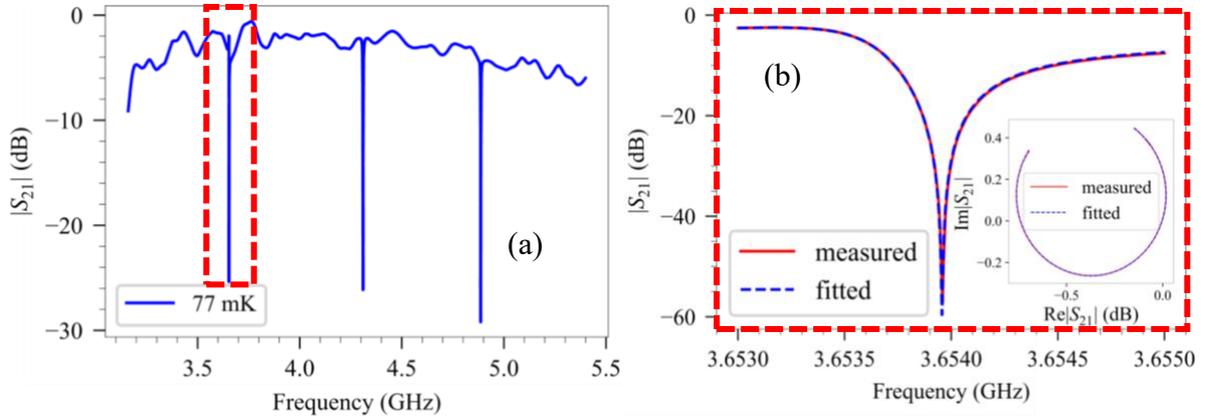

Figure 3. (a) Frequency spectrum of the microwave superconducting Ta CPW on silicon chip measured at $T = 77$ mK. (b) The magnitude and (inset) circle fit of the measured transmitted signal for $f_r = 3.6539$ GHz at high power (VNA power $= 0$ dBm, corresponding to $< n_{ph} > \approx 5.52 \times 10^5$ with $Q_l = 4872$ and $Q_c = 4897$ at $T = 77$ mK.

The photon number inside the $\frac{\lambda}{4}$ resonator can be determined by [46, 62]:

$$P_{in} = P_{trans} + P_{ref} + P_{abs}$$

$$P_{ref} = P_{in} (|S_{11}|^2)$$

$$P_{trans} = P_{in} (|S_{21}|^2)$$

$$P_{abs} = P_{in} (1 - |S_{21}|^2 - |S_{11}|^2)$$

$$<n_{ph}> = \frac{2Q_c}{\omega_0} \left(\frac{Q_i}{Q_i + Q_c}\right)^2 \frac{P_{in}}{\hbar \omega_0}$$

Where $P_{in}$ is the input power at the resonator, calculated as ($P_{in} = P_{VNA} + P_{fridge\ att} + P_{RT\ att}$), which $P_{fridge\ att}$ is the attenuation inside the fridge and $P_{RT\ att}$ is the room temperature attenuations, $P_{ref}$ is the reflected power, $P_{trans}$ is the transmitted power and $P_{abs}$ is the power absorbed by the resonator. In Fig.4, the relationship between resonators' $Q_i$ and the average photon number $<n_{ph}>$ is illustrated. For $f_r = 3.654$ GHz, the $Q_i$ is about $\sim 2.7 \times 10^5$ and $1.1 \times 10^6$ at single photon and many-photon regimes ($<n_{ph}> = 6.93 \times 10^5$), respectively. The variation in $Q_i$ is attributed to the TLS loss mechanism. TLS loss is temperature and power-dependent, reaching its maximum at low (milli-kelvin) temperature and low power (single-photon) regime. As power and temperature increase, the TLS loss decreases, which is known as the TLS saturation effect. Thus, an increase in power, leads to a decrease in the TLS loss, resulting the increasing $Q_i$. The resonance frequencies $f_r$, $Q_i$ at single photon and high power regimes, $Q_c$, $Q_l$ and $\frac{1}{Q_{TLS}^0}$ for all Ta thicknesses of 40 nm, 80 nm and 100 nm are given in Table 1. It can be seen that 100 nm Ta has the highest $Q_i$ at high power regimes.

Table 1. Comparison of superconducting microwave CPWs with different thicknesses of Ta films.

| Thickness | $f_r$ | $Q_i$ at single photon | $Q_i$ at high power | $\frac{1}{Q_{TLS}^0}$ | $Q_c$ | $Q_l$ |
|---|---|---|---|---|---|---|
| 40 nm | 3.654 GHz | $2.7 \times 10^5$ | $1.076 \times 10^6$ | $6.11 \times 10^{-6}$ | 4897 | 4872 |
| 40 nm | 4.31 GHz | $1.6 \times 10^5$ | $7.3 \times 10^5$ | $7.3 \times 10^{-6}$ | 1045 | 1043 |
| 40 nm | 4.88 GHz | $2 \times 10^5$ | $9.1 \times 10^5$ | $8.4 \times 10^{-6}$ | 1455 | 1453 |
| 80 nm | 4.2 GHz | $1.1 \times 10^5$ | $5.5 \times 10^5$ | $5.4 \times 10^{-6}$ | 3459 | 3438 |
| 80 nm | 4.9 GHz | $1.5 \times 10^5$ | $1.82 \times 10^5$ | $1.4 \times 10^{-6}$ | 1415 | 1405 |
| 80 nm | 5.64 GHz | $5 \times 10^5$ | $1.85 \times 10^6$ | $1.5 \times 10^{-6}$ | 2664 | 2660 |
| 100 nm | 4.3 GHz | $2 \times 10^5$ | $8 \times 10^5$ | $5.35 \times 10^{-6}$ | 5280 | 5240 |
| 100 nm | 5.1 GHz | $1.65 \times 10^5$ | $2.85 \times 10^5$ | $2.73 \times 10^{-6}$ | 863 | 860 |
| 100 nm | 5.8 GHz | $4.5 \times 10^5$ | $3.6 \times 10^6$ | $1.5 \times 10^{-6}$ | 1191 | 1190 |

The common TLS model is defined by [63]:

$$\delta_{TLS}(T,P) = \frac{1}{Q_{TLS}} = \frac{1}{Q_{TLS}^0} \frac{\tanh\left(\frac{hf_r}{2k_BT}\right)}{\sqrt{1+\left(\frac{n_{ph}}{n_c}\right)^\beta}} \qquad (4)$$

The TLS loss at zero power and temperature ($<n_{ph}> = 0$ and $T = 0$) is given by $\frac{1}{Q_{TLS}^0} = \delta^0{}_{TLS}$. Here, $n_c$ represents the critical photon number, $<n_{ph}>$ is the average photon number, and $\beta$ is known to be design-dependent [64, 65].

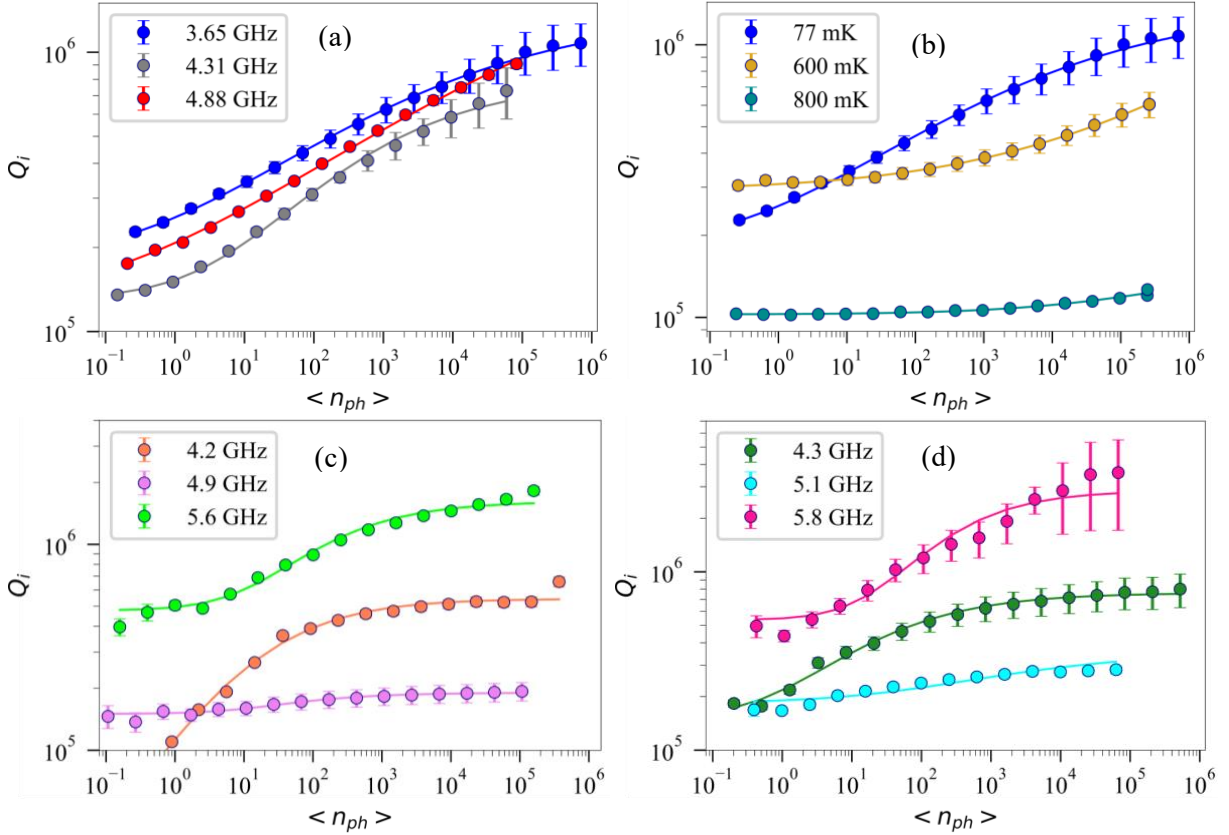

Figure 4. (a) Internal quality factor ($Q_i$) of three CPW resonators on silicon as a function of average photon numbers $<n_{ph}>$, the scatter plots are for measurement data, solid lines are fitted data based on Eq. (4) and error bars are depicted with caps at the top and bottom of each data point for (a) 40 nm Ta at $T = 77$ mK. (b) $f_r = 3.654$ at three different temperatures. (c) 80 nm Ta at $T = 44$ mK and (d) 100 nm Ta at $T = 40$ mK.

We fitted the measured data with Eq. (4) for all resonance frequencies and thicknesses at the base temperature $T = 77$ mK for 40 nm Ta (Fig.4 (a)), for $f_r = 3.654$ GHz at three different temperatures (Fig.4 (b)), for 80 nm Ta (Fig.4 (c)) and 100 nm Ta (Fig.4 (d)). For $f_r = 3.654$ GHz (40 nm thickness) at $T = 77$ mK, we obtained $\frac{1}{Q_{TLS}^0} = 6.11 \times 10^{-6}$ and $\beta = 0.44$ which indicates the strength of TLS saturation with power [66]. The total loss is the sum of TLS loss, quasi-particle loss and $\delta_{other}$ (e.g. other losses such as radiation loss, and the finite surface resistance of superconductors loss)[67]:

$$\frac{1}{Q_i} = \delta_i = \delta_{TLS}(T, P) + \delta_{qp}(T) + \delta_{other} \tag{5}$$

In Fig.4 (b), the comparison of $Q_i$ as a function of photon number at three different temperatures $T = 77$ mK and $T = 600$ mK and $T = 800$ mK for $f_r = 3.654$ GHz is presented. It is obvious that the sample at $T = 77$ mK has a higher $Q_i$ compared to $T = 800$ mK due to loss of system which was discussed earlier. The coupling quality factor ($Q_c$) determines the energy exchange between a resonator and its external environment such as measurement setup. Since $Q_c$ is design-dependent, it is theoretically expected to be constant for a fixed design at all power levels. As a result, when the $Q_c$ is constant and independent of power during the measurement, this consistency shows the stable coupling with minimal influence from the noise or external interference from measurement setup. In contrast, variable $Q_c$, indicating that measurement setup or external conditions affect the resonator's performance, which gives rise to instability or noise[68].

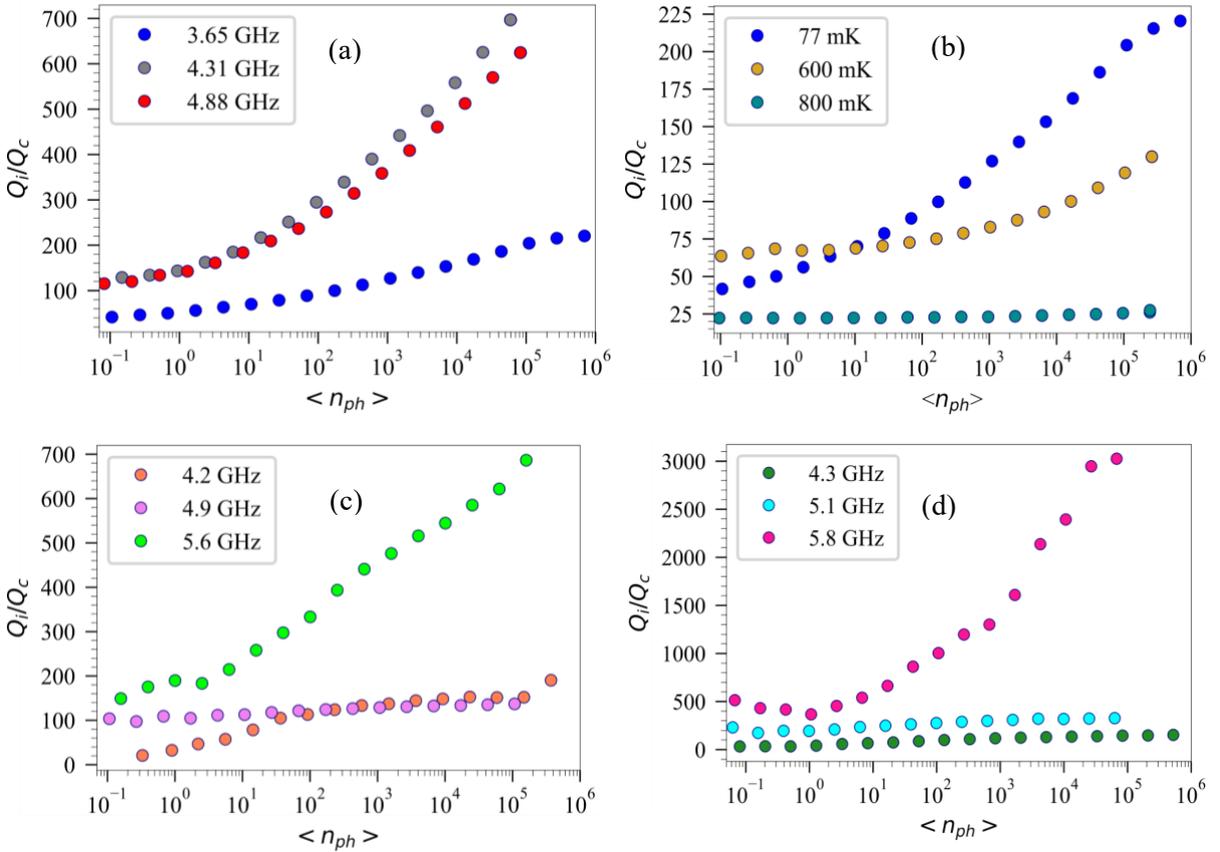

Figure 5. ($Q_i/Q_c$) of three CPW resonators on silicon as a function of photon number for (a) 40 nm Ta, at $T = 77$ mK. (b) 40 nm Ta for $f_r = 3.654$ at three different temperatures. (c) 80 nm Ta at $T = 44$ mK, and (d) 100 nm Ta at $T = 40$ mK.

In the overcoupled regime ($Q_i \gg Q_c$), due to the coupling of the larger fraction of the resonator's energy to the external circuit, makes the resonator more susceptible to external noise. But a constant $Q_c$, ensures that coupling strength is stable and measurement results reflect the true resonator behaviour without interference from external variables. Figure 5. shows the ratio of $\frac{Q_i}{Q_c}$ for all thicknesses at their resonance frequencies as a function of photon number [6, 52, 57, 69]. Since the $Q_c$ is constant, we can see a linear increase in $\frac{Q_i}{Q_c}$ by increasing the photon number (power).

In this section, the temperature dependence of $Q_i$ for 40 nm Ta is discussed. As noted in previous sections, the quasi-particle effect is an additional source of temperature-dependent losses. As the temperature increases, the TLS loss becomes negligible, and the quasi-particle becomes the dominant factor. Increasing the temperature results in a rising density of quasi-particles ($n_{qp}(T)$) leading to a reduction in $Q_i$. Consequently, $n_{qp}(T)$ determines the losses in the superconducting resonators and the loss model for quasi-particles can be defined by [70]:

$$\delta_{qp}(T) = \frac{1}{Q_{qp}} = \frac{\alpha}{\pi} \sqrt{\frac{2\Delta}{hf_r} \frac{n_{qp}(T)}{D(E_F)\Delta}} \tag{6}$$

where $\alpha$ is the ratio between the kinetic and total inductance of the resonator, $\Delta$ is the superconducting energy gap, $D(E_F)$ is the density of states at the Fermi level and $n_{qp}(T)$ is the density of a quasi-particle. Additionally, the contribution of quasi-particles can be explained by the Mattis-Bardeen theory [64, 71]:

$$\delta_{qp}(T) = \frac{2\gamma}{\pi} \frac{e^{-\varsigma}\sinh(\xi)K_0(\xi)}{1-e^{-\varsigma}(\sqrt{\frac{2\pi}{\varsigma}}-2e^{-\xi}I_0(\xi))} \tag{7}$$

Where $\gamma$ is the ratio of kinetic inductance to the total inductance of the conductor. $I_0$ and $K_0$ are modified Bessel functions of the first and second kind, $\varsigma = \frac{\Delta}{k_B T}$ and $\xi = \frac{\hbar\omega}{2k_B T}$.

Figure 6. (a) illustrates the dependence of $Q_i$ versus temperature for three resonators in the 40 nm Ta superconducting circuit. As previously discussed, in addition to TLS loss, the quasi-particle loss is also temperature-dependent. As shown in Fig. 6 (a), at low temperatures, the main factor altering $Q_i$ is the high TLS loss, resulting in a lower $Q_i$. As temperature increases, the TLS loss decreases, leading to an increase in $Q_i$. However, by increasing temperature to around $T$= 550 mK and beyond, quasi-particle losses become dominant, causing a subsequent

decrease in $Q_i$. Figure 6. (b) shows the measured and calculated resonance frequency shift ($\Delta f$) as a function of temperature. Here, $\Delta f = f_r(T) - f_r(T = 77 \text{ mK})$ and $\frac{\Delta f}{f_r} = \frac{f_r(T) - f_r(T = 77 \text{ mK})}{f_r(T = 77 \text{ mK})}$. The red solid line shows the calculated data based on Eq. (6), where $\alpha$ is obtained from the ratio of kinetic inductance to the total inductance.

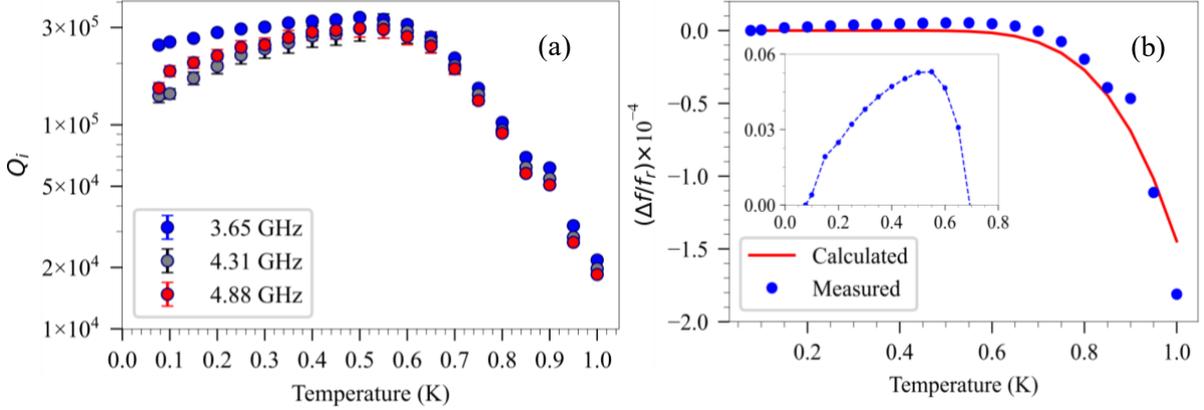

Figure 6. (a) Temperature dependence of the $Q_i$ for three resonators (40 nm Ta) at the single photon regime (error bars depicted with caps at the top and bottom of each data point). (b) The measured and calculated shift of resonance frequency as a function of temperature with the zoomed-in view (inset) at the single photon regime for $f_r = 3.654$ GHz (The dashed line at the inset is for the eye guide).

The inset in Fig. 6 (b) represents blueshift as the temperature increases from $T = 77$ mK to $T = 550$ mK, which is attributed to TLS loss which can be modelled by Eq. (8) [1]. Subsequently, a significant redshift is observed from $T = 550$ mK to $T = 1$ K. This redshift results from the increased density of quasi-particles at higher temperatures, leading to an increase in the kinetic inductance and a leftward shift of resonance frequency (redshift) as described by Eq. (9) [1]. Therefore, we can conclude that the total frequency shift is due to both TLS and quasi-particle losses $\Delta f \approx \Delta f_{TLS} + \Delta f_{qp}$. At low temperatures, the shift of resonance frequency is mainly due to TLS loss ($\Delta f \approx \Delta f_{TLS}$) and at high temperatures, it is mostly due to the quasi-particle effect ($\Delta f \approx \Delta f_{qp}$) which increases the kinetic inductance and as a result, shifts of resonance frequency to the lower frequencies.

Figure 7. (a) illustrates the colour-coded measured amplitude as a function of frequency in the single photon regime at temperatures ranging from 77 mK to 1K for $f_r = 3.654$ GHz. As the temperature increases from $T = 77$ mK to $T = 1$ K, the resonance frequency shifts to lower frequencies and the amplitude diminishes after $T = 550$ mK. As a result, the measured quality

factor decreases. Figure 7. (b) shows the colour-coded representation of phase, mirroring the amplitude's behaviour with a shift towards lower resonance frequencies after $T = 550$ mK. As temperature increases, due to an increase in the density of quasi-particles, the kinetic inductance increases which leads to a shift of resonance frequency to the lower resonance frequencies (redshift).

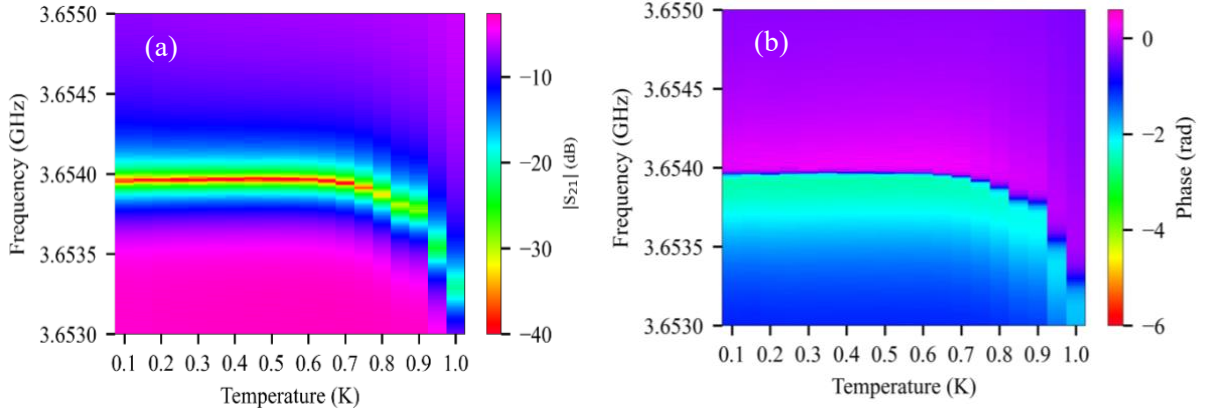

Figure 7. Color-map of (a) amplitude (b) phase of $S_{21}$ at different temperatures from $T = 77$ mK to $T = 1$ K at single photon regime for 40 nm Ta.

$$\Delta f_{TLS} = \frac{1}{Q_{TLS}^0} \frac{1}{\pi} \left[ R \left\{ \Psi \left( \frac{1}{2} - \frac{hf_r}{2\pi i k_B T} \right) \right\} - ln \frac{hf_r}{2\pi k_B T} \right] \quad (8)$$

where $\Psi$ is the digamma function.

$$\Delta f_{qp} = -\frac{1}{2} f_r \frac{\Delta L}{L} = -\frac{1}{2} \alpha f_r \frac{\Delta L_k}{L_k} = -\frac{1}{2} \alpha f_r \frac{\Delta}{k_B T} \frac{1}{\sinh \frac{\Delta}{k_B T}} \quad (9)$$

The comparison between previously reported superconducting microwave CPW resonators and this work is shown in Table 2. In works reported in references [31-33, 39], Ta films were sputtered on silicon substrates with a seed layer at room temperature, in contrast, in [35], heated silicon substrate without a seed layer was used. In works reported in [30, 36, 37], Ta was sputtered on the heated sapphire substrate without a buffer layer whereas in reference [38], a buffer layer was used. Our study features the thickness-dependent investigation of Ta thin films and focuses also on thin Ta film in a new approach, where the $Q_i$ observed between $1.6\text{-}5 \times 10^5$ at single photon to $0.73\text{-}3.6 \times 10^6$ at high power regimes indicating the production of high-quality Ta superconducting microwave circuits and their potential for applications in quantum technologies.

Table 2. Comparison of different approaches for the fabrication of Ta microwave CPW resonators.

| Ref | Substrate | Thickness of SC | Growth temperature | Type | Metal Preparation | T (mK) | $Q_i$ single photon | $Q_i$ high power |
|---|---|---|---|---|---|---|---|---|
| [33] | Si | 150 nm Ta/6 nm Nb seed layer | Room temperature | $\frac{\lambda}{4}$ | Sputter | - | - | ~ $10^6$ |
| [35] | Si | 100 nm | 400-450-500 °C | - | Sputter | 10 | 0.2-4.5 × $10^6$ | 55 × $10^6$ |
| [39] | Si | 100 nm Ta/5 nm Nb seed layer | Room temperature | $\frac{\lambda}{4}$ | - | 300 | < $10^5$ | ~ $10^5$ |
| [30] | Al$_2$O$_3$ | 150 nm | 550 °C | $\frac{\lambda}{4}$ | MBE | 10 | 0.9-1.3 × $10^6$ | 1.2-2.4 × $10^6$ |
| [31] | Si | 200 nm Ta/6 nm Nb seed layer | Room temperature | $\frac{\lambda}{4}$ | Sputter | 10 | - | 2 × $10^7$ |
| [36] | Al$_2$O$_3$ | 200 nm | 750 °C | $\frac{\lambda}{4}$ | Sputter | 17 | $10^5$-$10^7$ | $10^7$ - 2 × $10^8$ |
| [72] | Si, Al$_2$O$_3$, GaAs | 50 nm | Cryogenic | $\frac{\lambda}{4}$ | Low temperature MBE | 35 | 1.9 × $10^6$ for Si | > $10^7$ |
| [38] | Al$_2$O$_3$ | 300 nm/5 nm Nb seed layer | 500 °C | $\frac{\lambda}{4}$ | Sputter | 13 | 2.5 × $10^6$ | 27 × $10^6$ |
| [37] | Al$_2$O$_3$ | 200 nm | 400-500 °C | $\frac{\lambda}{4}$ | Sputter | 10 | 0.2-0.67 × $10^6$ | 0.35-27 × $10^6$ |
| **This work** | **Si** | 40 nmTa/6 nm Nb seed layer | **Room temperature** | $\frac{\lambda}{4}$ | **Sputter** | 77 | 1.6-2.7 × $10^5$ | 0.77-1.1 × $10^6$ |
| | **Si** | 80 nmTa/6 nm Nb seed layer | **Room temperature** | $\frac{\lambda}{4}$ | **Sputter** | 40 | 1.82-5 × $10^5$ | 55-1.85 × $10^6$ |
| | **Si** | 100 nmTa/6 nm Nb seed layer | **Room temperature** | $\frac{\lambda}{4}$ | **Sputter** | 40 | 1.65-4.5 × $10^6$ | 0.8-3.6 × $10^6$ |

In conclusion, we fabricated and characterised low-loss, high-quality superconducting microwave integrated circuits based on $\alpha$-Ta thin films of various thicknesses, utilizing the Nb seed layer on unheated high-resistivity silicon substrates. Silicon is the preferred choice for a wide range of electronic devices due to its compatibility with advanced silicon wafer technology and CMOS processing, which make it highly attractive for scalable quantum devices. The CPW resonators based on $\alpha$-Ta films showed an internal quality factor $Q_i$ exceeding $10^6$. Additionally, we investigated the effect of temperature and microwave power on $Q_i$, finding that TLS and quasi-particle losses as the two main factors influencing the resonator properties. Moreover, we characterized and calculated the critical temperature and kinetic inductance for different Ta films, observing an increase trend in $L_K$ by decreasing film thickness. Our approach holds potential for applications in quantum technologies, especially where high-$Q$, low loss with high kinetic inductance superconducting circuits compatible with semiconducting circuits are required. To further improve Ta superconducting circuit qualities, future research could focus on improving deposition methods, material purity, and interface engineering, which would increase their usefulness and scalability for realistic CMOS integration.

**Acknowledgement**

The authors thank the staff of the James Watt Nanofabrication Centre (JWNC) at the University of Glasgow. This work was supported in part by the Royal Society of Edinburgh, Royal Society, the EPSRC PRF-11-I-08, EPSRC EP/X025152/1, Innovate UK (QTools, grant No 79373 and FABU, grant No 50868); the EPSRC EP/T001062/1; and the FET Open initiative from the European Union's Horizon 2020 program under No 899561.